\documentclass[12pt, a4paper]{iopart}
\usepackage{iopams}
\usepackage[dvips]{graphicx}

\newcommand{\text}[1]{\mbox{\scriptsize #1}}

\begin{document}

\title{A class of vector identities relevant to the representation of the electric current density}

\author{M~Bornatici and O~Maj}

\address{Physics Department ``A.~Volta", University of Pavia, I-27100 Pavia, Italy}

\ead{maj@fisicavolta.unipv.it}

\begin{abstract}
A rigorous mathematical proof is given of a class of vector identities that provide a way to separate an arbitrary vector field (over a linear space) into the sum of a radial (i.e., pointing toward the radial unit vector) vector field, minus the divergence of a tensor plus the curl of an axial vector. Such a separation is applied to the representation of electric current densities yielding a specific form of the effective polarization and magnetization fields which is also discussed in some details. 
\end{abstract}



On dealing with the description of the electromagnetic response of material media, the issue of the representation of the induced current density appears to be crucial for the definition of the effective polarization and magnetization fields and, thus, for writing the macroscopic Maxwell's equations [Vinogradov and Aivazyan 1999, Vinogradov 2002, Agranovich \etal 2004, Bornatici and Maj 2005]. Specifically, one should find the effective polarization and magnetization fields, ${\bf P}_{\text{eff}}$ and ${\bf M}_{\text{eff}}$, as functionals of electric the current density ${\bf J}$ so that the basic relationship [Jackson 1999]
\begin{equation}
\label{1}
{\bf J}({\bf r},t) = \frac{\partial {\bf P}_{\text{eff}}({\bf r},t)}{\partial t} + c \nabla \times{\bf M}_{\text{eff}}({\bf r},t),
\end{equation}
is satisfied (Gaussian units are used). In particular, on accounting for the charge continuity equation, the representation \eref{1} yields the corresponding relationship for the charge density $\rho({\bf r},t)$, namely, $\rho({\bf r},t)=-\nabla\cdot{\bf P}_{\text{eff}}({\bf r},t)$, the required integration constant being properly chosen. 

It is readily verified that the effective fields ${\bf P}_{\text{eff}}$ and ${\bf M}_{\text{eff}}$ verifying equation \eref{1} for a given current density ${\bf J}$ are not uniquely determined and thus different forms of the effective polarization and magnetization fields are available, obtained on the basis of either the Helmholtz's theorem [Bornatici and Maj 2005] or vector identities proved by Vinogradov and Aivazyan [Vinogradov and Aivazyan 1999, Vinogradov 2002] and by Raab (quoted in [Vinogradov and Aivazyan 1999]).

In this letter, a rigorous mathematical proof is given of a class of identities which are satisfied by any (sufficiently regular) vector field over a linear space. When applied to the current density ${\bf J}({\bf r},t)$ regarded as a vector field over the three-dimensional ${\bf r}$-space (with time $t$ treated as a parameter), each one of such identities (labelled by an integer $n\geq 1$) can be cast in the form \eref{1} and, thus, yields a specific form of effective polarization and magnetization fields. The results reported by Vinogradov and Aivazyan [Vinogradov and Aivazian 1999, Vinogradov 2002] are found as particular cases for $n=1$ and $n=2$.  

With this aim, let us consider a generic vector field ${\bf R}({\bf x})=\big(R_{i}({\bf x}))_{i=1}^{N}$ with $N$ components over the $N$-dimensional (linear) space with ${\bf x}=(x_{1},\ldots,x_{N})$ (it is crucial that the dimension of the vector field is the same as that of the considered space) and assume that the field is sufficiently regular, i.e., differentiable up to a high-enough order. We claim that 
\begin{eqnarray}
\nonumber
\fl x_{i}x_{j_{1}}\cdots x_{j_{n-1}} \frac{\partial}{\partial x_{j_{1}}}\cdots \frac{\partial}{\partial x_{j_{n-1}}} \frac{\partial R_{k}}{\partial x_{k}} \\ 
\label{A1}
=\sum_{\ell=0}^{n-1} (-1)^{\ell} {n-1 \choose \ell} \frac{(N+n-1)!}{(N+n-\ell-1)!}
\left\{U^{(n-\ell)}_{i} -(n-\ell)q_{i}^{(n-\ell-1)} \right\},
\end{eqnarray}
where repeated indices are summed, $n\geq 1$ is an integer, ${n-1 \choose \ell}$ denotes the binomial coefficient and the following quantities have been defined as functionals of ${\bf R}({\bf x})$, {\it viz.},
\numparts
\begin{equation}
\label{A2a}
U^{(n)}_{i}({\bf x},{\bf R}({\bf x}))=\frac{\partial^{n}}{\partial x_{k_{1}}\cdots\partial x_{k_{n}}} \Big( x_{i}x_{k_{1}}\cdots x_{k_{n-1}} R_{k_{n}}({\bf x}) \Big), \qquad  (n\geq1),
\end{equation}
and
\begin{equation}
\label{A2b}
\eqalign{
q^{(0)}_{i}({\bf x},{\bf R}({\bf x})) = R_{i}({\bf x}), \\
q_{i}^{(n)}({\bf x}, {\bf R}({\bf x}))=\frac{\partial^{n} }{\partial x_{k_{1}}\cdots\partial x_{k_{n}}}\  q_{i k_{1}\cdots k_{n}}({\bf x},{\bf R}({\bf x})), \qquad\qquad\quad (n\geq1), 
}
\end{equation}
\endnumparts
with
\begin{eqnarray*}
q_{ik_{1}\cdots k_{n}}({\bf x},{\bf R}({\bf x})) &= x_{(i}x_{k_{1}}\cdots x_{k_{n-1}}R_{k_{n})}({\bf x}) \\
&\equiv \frac{1}{(n+1)!}\sum_{\sigma} x_{\sigma(i)}x_{\sigma(k_{1})}\cdots x_{\sigma(k_{n-1})} R_{\sigma(k_{n})}({\bf x})
\end{eqnarray*}
the symmetric part of the tensor $x_{i}x_{k_{1}}\cdots x_{k_{n-1}}R_{k_{n}}({\bf x})$, the index $\sigma$ running over the permutations of the $n+1$ indices $\{i,k_{1},\ldots, k_{n}\}$.
For the case $n=1$, equation (\ref{A1}) reduces to
\begin{equation*}
x_{i}\frac{\partial R_{k}}{\partial x_{k}} = \frac{\partial(x_{i}R_{k})}{\partial x_{k}} -R_{i},
\end{equation*}
which can be readily verified. Hence, let us prove the generic case by induction, that is, we need to prove that, if (\ref{A1}) is true for $n$, then it is true for $n+1$. With this aim, it is convenient to consider the identity
\begin{eqnarray*}
\fl x_{i}x_{j_{1}}\cdots x_{j_{n}}\frac{\partial}{\partial x_{j_{1}}}\cdots \frac{\partial}{\partial x_{j_{n}}} \frac{\partial R_{k}}{\partial x_{k}} = \frac{\partial}{\partial x_{j}} & \Big( x_{j} x_{i}x_{j_{1}}\cdots x_{j_{n-1}} \frac{\partial}{\partial x_{j_{1}}}\cdots\frac{\partial}{\partial x_{j_{n-1}}} \frac{\partial R_{k}}{\partial x_{k}} \Big)  \\
&-(N+n)\ x_{i}x_{j_{1}}\cdots x_{j_{n-1}} \frac{\partial}{\partial x_{j_{1}}}\cdots\frac{\partial}{\partial x_{j_{n-1}}} \frac{\partial R_{k}}{\partial x_{k}},
\end{eqnarray*}
which, on inserting equation (\ref{A1}), yields

\begin{eqnarray}
\nonumber
\fl x_{i}x_{j_{1}}\cdots x_{j_{n}}\frac{\partial}{\partial x_{j_{1}}}\cdots \frac{\partial}{\partial x_{j_{n}}} \frac{\partial R_{k}}{\partial x_{k}} \\
\nonumber
\fl\quad\quad\ =\sum_{\ell=0}^{n-1} (-1)^{\ell} {n-1 \choose \ell} \frac{(N+n-1)!}{(N+n-\ell -1)!} \frac{\partial}{\partial x_{j}}\left[x_{j}\Big( U^{(n-\ell)}_{i} -(n-\ell) q^{(n-\ell -1)}_{i} \Big)\right]  \\
\label{A3}
- \sum_{\ell=0}^{n-1} (-1)^{\ell} {n-1 \choose \ell} \frac{(N+n)!}{(N+n-\ell-1)!}\ \left[U^{(n-\ell)}_{i} - (n-\ell) q_{i}^{(n-\ell -1)} \right].
\end{eqnarray}  
Moreover one has
\begin{equation}
\label{A4}
\eqalign{
\frac{\partial}{\partial x_{j}}\Big(x_{j} U^{(n)}_{i} \Big) = U^{(n+1)}_{i} - nU^{(n)}_{i},\\
\frac{\partial}{\partial x_{j}}\Big(x_{j} q^{(n)}_{i} \Big) = \frac{n+2}{n+1} q_{i}^{(n+1)} -\frac{1}{n+1} U^{(n+1)}_{i} - nq_{i}^{(n)}, 
}
\end{equation}
where the identity
\begin{equation}
\label{A5}
q_{i}^{(n)} = \frac{1}{n+1} \frac{\partial^{n}}{\partial x_{k_{1}} \cdots \partial x_{k_{n}}} \Big( R_{i}x_{k_{1}}\cdots x_{k_{n}} + n\ x_{i}x_{k_{1}}\cdots x_{k_{n-1}}R_{k_{n}}  \Big)
\end{equation}
has been used. Equation (\ref{A3}), along with (\ref{A4}), reduces to (\ref{A1}) with $n \rightarrow n+1$, hence one can conclude that the identity (\ref{A1}) holds true for any integer $n\geq 1$.

By inspection of (\ref{A1}) one should note that it is convenient to separate the symmetric part of the tensor $x_{i}x_{k_{1}}\cdots x_{k_{n-1}}R_{k_{n}}$ in the quantity $U_{i}^{(n)}$, namely, (cf. equation (\ref{A2a}))  
\begin{eqnarray}
\nonumber
\fl U^{(n)}_{i} ({\bf x},{\bf R}({\bf x}))=\frac{\partial^{n}}{\partial x_{k_{1}}\cdots\partial x_{k_{n}}} \Big( x_{i}x_{k_{1}}\cdots x_{k_{n-1}} R_{k_{n}}({\bf x}) \Big)\\
\nonumber
\fl = \frac{\partial^{n}}{\partial x_{k_{1}}\cdots \partial x_{k_{n}}}\left[ \frac{1}{n+1}\Big(x_{i}x_{k_{1}}\cdots x_{k_{n-1}}R_{k_{n}} + x_{i}x_{k_{1}}\cdots R_{k_{n-1}}x_{k_{n}} + \cdots + R_{i}x_{k_{1}}\cdots x_{k_{n}} \Big)  \right.\\
\nonumber
\fl \quad \left. + \frac{n}{n+1} x_{i}x_{k_{1}}\cdots x_{k_{n-1}}R_{k_{n}} - \frac{1}{n+1}\Big( x_{i}x_{k_{1}}\cdots R_{k_{n-1}}x_{k_{n}} + \cdots + R_{i}x_{k_{1}}\cdots x_{k_{n}} \Big) \right]\\
\nonumber
\fl = q_{i}^{(n)} + \frac{\partial^{n}}{\partial x_{k_{1}}\cdots \partial x_{k_{n}} } \left[ \frac{1}{n+1}\big( 
x_{i}R_{k_{n}} - x_{k_{n}}R_{i} \big)x_{k_{1}}\cdots x_{k_{n-1}}\right]\\
\label{A6}
\fl = q_{i}^{(n)} + \varepsilon_{ijk}\frac{\partial}{\partial x_{j}} m_{k}^{(n)},
\end{eqnarray}
where the vector $q_{i}^{(n)}=q_{i}^{(n)}({\bf x},{\bf R}({\bf x}))$ is defined in equations \eref{A2b} and the axial vector $m_{k}^{(n)}=m_{k}^{(n)}({\bf x},{\bf R}({\bf x}))$ is given by
\numparts
\begin{eqnarray}
\label{A7a}
m_{k}^{(n)}({\bf x},{\bf R}({\bf x})) = \frac{\partial^{n-1}}{\partial x_{k_1}\cdots\partial x_{k_{n-1}}}\ m_{kk_{1}\cdots k_{n-1}}({\bf x},{\bf R}({\bf x})),\quad\quad (n\geq 1),\\
\nonumber
\fl \mbox{with}\\
\label{A7b}
m_{kk_{1}\cdots k_{n-1}}({\bf x},{\bf R}({\bf x}))= \frac{1}{n+1}\Big(\varepsilon_{kab}x_{a}R_{b}({\bf x})\Big)x_{k_{1}}\cdots x_{k_{n-1}}.
\end{eqnarray}
\endnumparts
In equation (\ref{A6}), the identities \eref{A5} and $\varepsilon_{ijk}\varepsilon_{kab} = \delta_{ia}\delta_{jb} - \delta_{ib}\delta_{ja}$ have been accounted for, $\varepsilon_{ijk}$ and $\delta_{ij}$ being, respectively, the completely anti-symmetric permutation symbol and the Kronecker delta. On substituting (\ref{A6}) into (\ref{A1}) and grouping together the terms of the same form, one has
\begin{eqnarray}
\nonumber
\fl x_{i}x_{j_{1}}\cdots x_{j_{n-1}}\frac{\partial}{\partial x_{j_{1}}}\cdots \frac{\partial}{\partial x_{j_{n-1}}}\frac{\partial R_{k}}{\partial x_{k}} = q_{i}^{(n)} \\
\nonumber
\fl\qquad + \sum_{\ell=1}^{n-1} (-1)^{\ell} \frac{(N+n-1)!}{(N+n-\ell-1)!} {n-1 \choose \ell-1}\left[ \frac{n}{\ell}-\frac{N-1}{N+n-\ell} \right]\ q_{i}^{(n-\ell)}\\
\label{A8}
\fl\qquad + \varepsilon_{ijk}\frac{\partial}{\partial x_{j}} \left[ m_{k}^{(n)} + \sum_{\ell=1}^{n-1} (-1)^{\ell} {n-1 \choose \ell}\ m_{k}^{(n-\ell)} \right] + (-1)^{n}\frac{(N+n-1)!}{N!}R_{i},
\end{eqnarray} 
where equations (\ref{A2b}) for $n=0$ has been used. Equation (\ref{A8}) can be solved for $R_{i}$ as a function of its own derivatives. To this end it is convenient to change the dummy index $\ell$ into $n-\ell$ with the result that, for any integer $n\geq 1$,
\begin{equation}
\label{new13}
{\bf R} = {\bf X}^{(n)} -\nabla\cdot\mathsf{T}^{(n)} + \nabla\times{\bf Y}^{(n)},
\end{equation}
where the fields ${\bf X}^{(n)}=(X^{(n)}_{i})_{i=1}^{N}$, $\mathsf{T}^{(n)}=(T^{(n)}_{ik})_{i,k=1}^{N}$ and ${\bf Y}^{(n)}=(Y^{(n)}_{k})_{k=1}^{N}$ are functionals of ${\bf R}({\bf x})$ given by
\numparts
\begin{eqnarray}
\label{new14a}
\fl X^{(n)}_{i}({\bf x},{\bf R}({\bf x}))= (-1)^{n}\frac{N!}{(N+n-1)!} x_{i}x_{j_{1}}\cdots x_{j_{n-1}}\frac{\partial}{\partial x_{j_{1}}}\cdots \frac{\partial}{\partial x_{j_{n-1}}}\frac{\partial R_{k}({\bf x})}{\partial x_{k}}, \\
\nonumber
\fl T^{(n)}_{ik}({\bf x},{\bf R}({\bf x}))=  \sum_{\ell=1}^{n} (-1)^{\ell} \frac{N!}{(N+\ell-1)!} {n \choose \ell}  \left[1-\frac{(n-\ell)(N-1)}{n(N+\ell)}\right]\\
\label{new14b}
\qquad\qquad \times \frac{\partial^{\ell-1}}{\partial x_{k_{1}}\cdots\partial x_{k_{\ell-1}}}\ q_{ikk_{1}\cdots k_{\ell-1}}({\bf x},{\bf R}({\bf x})), \\
\label{new14c}
\fl Y^{(n)}_{k}({\bf x},{\bf R}({\bf x}))= \sum_{\ell=1}^{n} (-1)^{\ell-1} \frac{N!}{(N+\ell-1)!} {n \choose \ell}\frac{\ell}{n}\frac{\partial^{\ell-1}}{\partial x_{k_{1}}\cdots \partial x_{k_{\ell-1}}} m_{kk_{1}\cdots k_{\ell-1}}({\bf x},{\bf R}({\bf x})).
\end{eqnarray}
\endnumparts

Equation \eref{new13} constitutes the main result of this paper according to which a generic vector field ${\bf R}({\bf x})$ can be decomposed into a radial vector ${\bf X}^{(n)}$, pointing toward the radial unit vector $\hat{\bf x}={\bf x}/|{\bf x}|$ (with $|{\bf x}|^{2}=\sum_{i} (x_{i})^{2}$), the divergence of a tensor $\mathsf{T}^{(n)}$ and the curl of an axial vector ${\bf Y}^{(n)}$. It is worth noting that for every integer $n\geq 1$, there is a set of fields ${\bf X}^{(n)}$, $\mathsf{T}^{(n)}$ and ${\bf Y}^{(n)}$ which satisfy identically equation \eref{new13}, the right-hand-side of \eref{new13} being, indeed, independent on $n$.  

The identity (\ref{new13}) can be readily applied to represent the current density ${\bf J}({\bf r},t)$ over the three-dimensional ${\bf r}$-space, the time $t$ being treated as a parameter. Specifically, for fixed time $t$, one can set ${\bf J}({\bf r},t)={\bf R}({\bf x})$, with ${\bf r}={\bf x}=(x_{1},x_{2},x_{3})$ and $N=3$, and, for any integer $n\geq 1$, define the effective polarization, ${\bf P}^{(n)}_{\text{eff}}$, and magnetization, ${\bf M}_{\text{eff}}^{(n)}$, by
\begin{equation*}
{\bf X}^{(n)}({\bf r},{\bf J}({\bf r},t)) - \nabla\cdot\mathsf{T}^{(n)}({\bf r},{\bf J}({\bf r},t)) = \frac{\partial {\bf P}^{(n)}_{\text{eff}}({\bf r},t)}{\partial t},
\end{equation*}
and
\begin{equation*}
{\bf Y}^{(n)}({\bf r},{\bf J}({\bf r},t))=  c\ {\bf M}^{(n)}_{\text{eff}}({\bf r},{\bf J}({\bf r},t)),
\end{equation*}
respectively, so that equation \eref{new13} amounts to the representation \eref{1} for the electric current density ${\bf J}({\bf r},t)$.

On accounting for equations \eref{new14a}-\eref{new14c} explicitly, the effective fields ${\bf P}^{(n)}_{\text{eff}}$ and ${\bf M}^{(n)}_{\text{eff}}$ can be expressed in a form which, formally, resembles the multipole expansion [Agranovich \etal 2004], namely,
\numparts
\begin{eqnarray}
\label{new16a}     
\fl\qquad P^{n}_{\text{eff}\ i}({\bf r},t) = P^{(n)}_{i}\big({\bf r},{\bf J}({\bf r},t)\big) - \frac{\partial }{\partial x_{k}}\sum_{\ell=1}^{n} \frac{\partial^{\ell-1}}{\partial x_{k_{1}}\cdots \partial x_{k_{\ell-1}}}\ Q^{(n,\ell)}_{ikk_{1}\cdots k_{\ell-1}}\big({\bf r},{\bf J}({\bf r},t)\big), \\
\label{new16b}
\fl\qquad M^{(n)}_{\text{eff}\ i}({\bf r},t)=\sum_{\ell=1}^{n} (-1)^{\ell-1} \frac{\partial^{\ell-1}}{\partial x_{k_{1}}\cdots \partial x_{k_{\ell-1}}}\ M^{(n,\ell)}_{ik_{1}\cdots k_{\ell-1}}\big({\bf r},{\bf J}({\bf r},t)\big),
\end{eqnarray}
\endnumparts
where $P_{i}^{(n)}$, $Q^{(n,\ell)}_{ikk_{1}\cdots k_{\ell-1}}$ and $M^{(n,\ell)}_{kk_{1}\cdots k_{\ell-1}}$play the role of electric dipole, electric quadrupole and magnetic multipole moments, respectively, and are functionals of the induced current density ${\bf J}({\bf r},t)$ given by
\numparts
\begin{eqnarray}
\label{new17a}
\fl \frac{\partial P_{i}^{(n)}}{\partial t}\big({\bf r},{\bf J}({\bf r},t)\big) =(-1)^{(n)}\frac{6}{(n+2)!} x_{i}x_{j_{1}}\cdots x_{j_{n-1}}\frac{\partial}{\partial x_{j_{1}}}\cdots \frac{\partial}{\partial x_{j_{n-1}}}\frac{\partial J_{k}({\bf r},t)}{\partial x_{k}},\\
\label{new17b}
\fl \frac{\partial Q^{(n,\ell)}_{ikk_{1}\cdots k_{\ell-1}}}{\partial t} \big({\bf r},{\bf J}({\bf r},t)\big) = (-1)^{\ell}\frac{6}{(\ell+2)!} {n \choose \ell}\left[1-\frac{2(n-\ell)}{n(\ell+3)}\right] x_{(i}x_{k_{1}}\cdots x_{k_{\ell-1}}J_{k)}({\bf r},t),\\
\label{new17c}
\fl M^{(n,\ell)}_{ik_{1}\cdots k_{\ell-1}} \big({\bf r},{\bf J}({\bf r},t)\big) = \frac{6}{c(\ell+2)!} {n \choose \ell}\frac{\ell}{n(\ell+1)}\big(\varepsilon_{ijk}x_{j}J_{k}({\bf r},t)\big)x_{k_{1}}\cdots x_{k_{\ell-1}}.
\end{eqnarray}
\endnumparts
Equation \eref{1} along with the effective fields (12) takes the form 
\begin{eqnarray}
\nonumber
\fl J_{i}= \frac{\partial P_{i}^{(n)}}{\partial t} - \frac{\partial}{\partial t}\frac{\partial}{\partial x_{k}}\sum_{\ell=1}^{n} \frac{\partial^{\ell-1}}{\partial x_{k_{1}}\cdots \partial x_{k_{\ell-1}}}Q^{(n,\ell)}_{ikk_{1}\cdots k_{\ell-1}} \\
\label{new14}
\qquad\qquad\qquad+ c\ \varepsilon_{ijk}\frac{\partial}{\partial x_{j}}\sum_{\ell=1}^{n}(-1)^{\ell-1} \frac{\partial^{\ell-1}}{\partial x_{k_{1}}\cdots \partial x_{k_{\ell-1}}} M^{(n,\ell)}_{kk_{1}\cdots k_{\ell-1}},
\end{eqnarray}
which exhibits the same mathematical structure as a multipole expansion truncated to the $n$-th order, even though, with the ``multipoles'' given by equations \eref{new17a}-\eref{new17c}, it is an exact identity. 
 
For the specific case $n=1$, equation (\ref{new14}) reduces to
\begin{equation}
\label{A12}
J_{i} = -x_{i}\frac{\partial J_{k}}{\partial x_{k}} - \frac{\partial}{\partial x_{k}}\left[-\frac{1}{2}\big(x_{k}J_{i} + x_{i}J_{k}\big)\right] + c \varepsilon_{ijk}\frac{\partial}{\partial x_{j}}\left[ \frac{1}{2c} \varepsilon_{kmn}x_{m}J_{n}\right]
\end{equation}
which is the identity proved by Vinogradov and Aivazyan [Vinogradov and Aivazyan 1999].

For $n=2$, equation (\ref{new14}) reads
\begin{equation}
\label{A13}
\fl\qquad\qquad J_{i} = \frac{\partial P_{i}^{(2)}}{\partial t} - \frac{\partial}{\partial t}\frac{\partial Q_{ij}^{(2,1)}}{\partial x_{j}} - \frac{\partial}{\partial t} \frac{\partial^{2}Q^{(2,2)}_{ijk}}{\partial x_{j}\partial x_{k}} 
+ c\varepsilon_{ijk}\frac{\partial M_{k}^{(2,1)}}{\partial x_{j}} - c \varepsilon_{ijk}\frac{\partial^{2} M_{kl}^{(2,2)}}{\partial x_{j}\partial x_{l}}
\end{equation}
with 
\numparts
\begin{eqnarray}
\label{A14a}
&\frac{\partial P_{i}^{(2)}}{\partial t} \big({\bf r},{\bf J}({\bf r},t)\big) = \frac{1}{4} x_{i}x_{j}\frac{\partial}{\partial x_{j}}\frac{\partial J_{k}}{\partial x_{k}}\\
\label{A14b}
&\frac{\partial Q_{ij}^{(2,1)}}{\partial t} \big({\bf r},{\bf J}({\bf r},t)\big) = -\frac{3}{4}\big(x_{i}J_{j} + x_{j}J_{i}\big)\\
\label{A14c}
&\frac{\partial Q_{ijk}^{(2,2)}}{\partial t} \big({\bf r},{\bf J}({\bf r},t)\big) = \frac{1}{12}\big(x_{i}x_{j}J_{k} + x_{i}J_{j}x_{k} + J_{i}x_{j}x_{k}\big),\\
\label{A14d}
&M_{i}^{(2,1)} \big({\bf r},{\bf J}({\bf r},t)\big) = \frac{1}{2c} \varepsilon_{ijk}x_{j}J_{k},\\
\label{A14e}
&M_{il}^{(2,2)} \big({\bf r},{\bf J}({\bf r},t)\big) = \frac{1}{12c} \big(\varepsilon_{ijk}x_{j}J_{k}\big)x_{l}.
\end{eqnarray}
\endnumparts
Equation (\ref{A13}), along with (17), agrees with Raab's result quoted in [Vinogradov and Aivazyan 1999]. (The factor $1/2$ in the expression (\ref{A14d}) for ${\bf M}^{(2,1)}$ appears to be missing in the corresponding Raab's expression.)

With reference to (\ref{new17a}) and (\ref{new17b}), only the time-derivative of both the vector $P_{i}^{(n)}$ and the tensors $Q_{ikk_{1}\cdots k_{\ell-1}}^{(n,\ell)}$ is specified as functionals of the induced current density ${\bf J}({\bf r},t)$. The explicit form of both $P_{i}^{(n)}$ and $Q_{ikk_{1}\cdots k_{\ell-1}}^{(n,\ell)}$ is obtained on integrating with respect to time,
\begin{eqnarray}
\label{13a}
P_{i}^{(n)}\big({\bf r}, {\bf J}({\bf r},t)\big) = \int_{t_{0}}^{t}dt' \frac{\partial P_{i}^{(n)}}{\partial t'}\big({\bf r},{\bf J}({\bf r},t')\big) + \tilde{P}_{i}^{(n)}({\bf r}),\\
\label{13b}
Q_{ikk_{1}\cdots k_{\ell-1}}^{(n,\ell)}\big({\bf r}, {\bf J}({\bf r},t)\big) = \int_{t_{0}}^{t}dt' \frac{\partial Q_{ikk_{1}\cdots k_{\ell-1}}^{(n,\ell)}}{\partial t'}\big({\bf r},{\bf J}({\bf r},t')\big) + \tilde{Q}_{ikk_{1}\cdots k_{\ell-1}}^{(n,\ell)}({\bf r}),
\end{eqnarray}
where $t_{0}$ is a reference time (as it appears below, its presence is crucial for the treatment of static fields) and $\tilde{P}_{i}^{(n)}({\bf r})$ and $\tilde{Q}_{ikk_{1}\cdots k_{\ell-1}}^{(n,\ell)}({\bf r})$ are time-independent integration constants which should be selected in such a way that the representation in (\ref{1}) for the induced charge density $\rho({\bf r},t)$ is verified, i.e., $\nabla\cdot {\bf P}_{\text{eff}}^{(n)} ({\bf r},t) = -\rho({\bf r},t)$. 

In particular, the integration in (\ref{13a}) can be explicitly carried out on making use of (\ref{new17a}) along with the charge continuity equation, with the result that $P_{i}^{(n)}$ is a functional of the induced charge density,
\begin{eqnarray}
\nonumber
\fl P_{i}^{(n)}\big({\bf r},\rho({\bf r},t)\big) = (-1)^{n+1} \frac{6}{(n+2)!} x_{i}x_{j_{1}}\cdots x_{j_{n-1}}\frac{\partial}{\partial x_{j_{1}}}\cdots\frac{\partial}{\partial x_{j_{n-1}}} \Big( \rho({\bf r},t)- \rho({\bf r},t_{0}) \Big)\\
\label{14}
\qquad\quad+\ \tilde{P}_{i}^{(n)}({\bf r}).
\end{eqnarray}
For the specific case of \emph{fluctuating perturbations}, it is usually assumed that both $\rho({\bf r},t)$ and ${\bf J}({\bf r},t)$ tend to zero for $t \to -\infty$, so that one can set $t_{0}=-\infty$ in equations (\ref{13a}) and \eref{13b} as well as (\ref{14}), with both $\tilde{P}_{i}^{(n)}({\bf r})$ and $\tilde{Q}_{ikk_{1}\cdots k_{\ell-1}}^{(n,\ell)}({\bf r})$ being zero. On the other hand, for the case of \emph{static fields}, for which, in particular, $\rho({\bf r},t)=\rho({\bf r},t_{0})=\rho({\bf r})$, equation (\ref{14}) reduces to
\begin{equation}
\label{15a}
P^{(n)}_{i}\big({\bf r},\rho({\bf r})\big) = \tilde{P}^{(n)}_{i}({\bf r}),
\end{equation}
and equation (\ref{13b}) along with (\ref{new17b}) where ${\bf J}({\bf r},t)={\bf J}({\bf r})$ yields
\begin{eqnarray}
\nonumber
\fl
Q_{ikk_{1}\cdots k_{\ell-1}}^{(n,\ell)}\big({\bf r}, {\bf J}({\bf r})\big) = (-1)^{\ell}\frac{6(t-t_{0})}{(\ell+2)!}{n \choose \ell}\left[1-\frac{2(n-\ell)}{n(\ell+3)}\right]  x_{(i}x_{k_{1}}\cdots x_{k_{\ell-1}}J_{k)}({\bf r})\\
\label{15b}
\qquad\quad +\ \tilde{Q}_{ikk_{1}\cdots k_{\ell-1}}^{(n,\ell)}({\bf r}).
\end{eqnarray} 
One should note that equation (\ref{15b}) for the effective electric multipoles exhibits a time-dependence even though static fields are considered. As for the effective magnetization, the static case is simply given by (\ref{new17c}) with ${\bf J}({\bf r},t)$ replaced by ${\bf J}({\bf r})$.

Let us examine in some detail the effective fields (\ref{new16a}) and \eref{new16b} for the specific case of $n=1$ [Vinogradov and Aivazyan 1999]. With $n=1$, the effective polarization (\ref{new16a}) is
\begin{equation}
\label{16}
{\bf P}_{\text{eff}}^{(1)}({\bf r},t) = {\bf P}^{(1)}\big({\bf r},\rho({\bf r},t)\big) - \nabla\cdot \mathsf{Q}^{(1,1)}\big({\bf r},{\bf J}({\bf r},t)\big),
\end{equation}
with, from (\ref{14}),
\begin{equation}
\label{17new}
{\bf P}^{(1)}\big({\bf r},\rho({\bf r},t)\big)= 
\cases{
\rho({\bf r},t){\bf r},\\ \\
\tilde{\bf P}^{(1)}({\bf r}),}
\end{equation}
and, from (\ref{13b}) along with (\ref{new17b}),
\begin{equation} 
\label{18new}
\mathsf{Q}^{(1,1)}({\bf r},{\bf J}({\bf r},t)\big) = 
\cases{
-\frac{1}{2}\int_{-\infty}^{t}dt' \Big({\bf r}{\bf J}({\bf r},t') + {\bf J}({\bf r},t'){\bf r} \Big),\\ \\
-\frac{(t-t_{0})}{2} \Big({\bf r}{\bf J}({\bf r}) + {\bf J}({\bf r}){\bf r} \Big) + \tilde{\mathsf{Q}}^{(1,1)}({\bf r}),
}
\end{equation}
the upper (lower) entry of (\ref{17new}) and (\ref{18new}) referring to fluctuating (static) fields. On making use of the charge continuity equation, from (\ref{18new}) one gets
\begin{equation}
\label{19new}
\fl \nabla\cdot \mathsf{Q}^{(1,1)}\big({\bf r},{\bf J}({\bf r},t)\big) = 
\cases{
\frac{1}{2} \rho({\bf r},t){\bf r} - \int_{-\infty}^{t}dt' \Big( 2{\bf J}({\bf r},t') +\frac{1}{2} ({\bf r}\cdot\nabla){\bf J}({\bf r},t') \Big),\\ \\
-\frac{(t-t_{0})}{2} \Big(4{\bf J}({\bf r}) + ({\bf r}\cdot \nabla){\bf J}({\bf r}) \Big) + \nabla\cdot \tilde{\mathsf{Q}}^{(1,1)}({\bf r}).
}
\end{equation}
It is readily verified that the effective polarization (\ref{16}) along with (\ref{17new}) and (\ref{18new}) is such that $\nabla \cdot {\bf P}_{\text{eff}}^{(1)}({\bf r},t)=-\rho({\bf r},t)$ this result requiring that $\nabla\cdot\big( \tilde{\bf P}^{(1)}({\bf r}) - \nabla \cdot\tilde{\mathsf{Q}}^{(1,1)}({\bf r})\big)=-\rho({\bf r})$ for the case of static fields.

As for the magnetization (\ref{new16b}), for $n=1$, it is
\begin{equation}
\label{20}
{\bf M}_{\text{eff}}^{(1)}({\bf r},t)={\bf M}^{(1,1)}\big({\bf r},{\bf J}({\bf r},t)\big) = \frac{1}{2c} \big({\bf r}\times{\bf J}({\bf r},t)\big),
\end{equation}
which agrees with the conventional definition of \emph{magnetic moment density} associated to the current density ${\bf J}({\bf r},t)$ [Jackson 1999, Vinogradov and Aivazyan 1999].

\section*{References}
\begin{harvard}
\item[] Bornatici M and Maj O 2006 {\it Physica Scripta} {\bf 73}, 160
\item[] Jackson J D 1999 {\it Classical Electrodynamics} (New York: Wiley)
\item[] Vinogradov A P and Aivazyan A V 1999 {\it Phys. Rev. E} {\bf 60}, 987
\item[] Vinogradov A P 2002 {\it Physics Uspekhi} {\bf 45}, 331 
\end{harvard}

\end{document}